\documentclass[aps, prd, floatfix, nofootinbib, superscriptaddress, twocolumn]{revtex4-1}

\usepackage{latexsym}
\usepackage{amsmath}
\usepackage{amssymb}
\usepackage{amsfonts}

\usepackage[mathscr,scaled=1.15]{urwchancal}
\DeclareFontFamily{OT1}{pzc}{}
\DeclareFontShape{OT1}{pzc}{m}{it}%
{<-> s * [1.15] pzcmi7t}{}
\DeclareMathAlphabet{\mathpzc}{OT1}{pzc}{m}{it}

\usepackage{color}

\usepackage{supertabular}
\usepackage{placeins}
\usepackage{epsfig}
\usepackage{graphicx}

\definecolor{purple}{rgb}{0.5,0,0.5}
\definecolor{blue}{rgb}{0.0,0,0.9}
\definecolor{prdblue}{rgb}{0.133,0.118,0.498}
\usepackage[colorlinks=true, pdfstartview=FitV, linkcolor=prdblue, citecolor= prdblue, urlcolor=prdblue]{hyperref}

\begin{document}


\title{$\,$\\[-7ex]\hspace*{\fill}{\normalsize{\sf\emph{Preprint no}. NJU-INP 042/21}}\\[1ex]
Regarding the distribution of glue in the pion}

\author{Lei Chang}
\email[]{leichang@nankai.edu.cn}
\affiliation{School of Physics, Nankai University, Tianjin 300071, China}

\author{Craig D.~Roberts}
\email[]{cdroberts@nju.edu.cn}
\affiliation{School of Physics, Nanjing University, Nanjing, Jiangsu 210093, China}
\affiliation{Institute for Nonperturbative Physics, Nanjing University, Nanjing, Jiangsu 210093, China}

\date{2020 June 15}

\begin{abstract}
Understanding why the scale of emergent hadron mass is obvious in the proton but hidden in the pion may rest on mapping the distribution functions (DFs) of all partons within the pion and comparing them with those in the proton; and since glue provides binding in quantum chromodynamics, the glue DF could play a special role.  Producing reliable predictions for the proton's DFs is difficult because the proton is a three valence-body bound-state problem.  As sketched herein, the situation for the pion, a two valence-body problem, is much better, with continuum and lattice predictions for the valence-quark and glue DFs in agreement.  This beginning of theory alignment is timely because experimental facilities now either in operation or planning promise to realise the longstanding goal of providing pion \emph{targets}, thereby enabling precision experimental tests of rigorous theory predictions concerning Nature's most fundamental Nambu-Goldstone bosons.
%
%
\end{abstract}

\maketitle



\noindent\emph{1.$\;$Introduction}.\,---\,%
The proton was discovered in 1917 \cite{RutherfordIV}.  It is stable: looking back through the approximately 14-billion year history of the Universe, no proton in isolation has ever been seen to decay.  This is one of the most profound features of Nature; and for science it means, \emph{inter alia}, that protons can readily be used as targets or accelerated as probes for the exploration of other material \cite{Cockroft:1932I}.  These properties have been very profitably exploited, enabling the discovery of quarks \cite{Taylor:1991ew, Kendall:1991np, Friedman:1991nq} and fostering the development of quantum chromodynamics (QCD) as the theory of strong interactions \cite{Marciano:1979wa}.  

When discussing proton structure, parton distribution functions (DFs) are often the focus \cite{Brock:1993sz}.  Each DF, ${\mathpzc p}(x;\zeta)$, describes the number density of a given parton type, ${\mathpzc p}$ = valence-quark (${\mathpzc q}$), sea-quark (${\mathpzc S}$), or glue ($\mathpzc g$), within the proton as a function of the light-front fraction, $x\in (0,1)$, of the proton's momentum that it carries at a resolving scale, $\zeta$.  As with many quantities in quantum field theory, like the coupling and masses of elementary fields, these densities depend on the energy scale with which they are probed.  Crucially, under certain kinematic conditions \cite{Brock:1993sz}, such as those prevailing in deep inelastic scattering (DIS) or Drell-Yan (DY) reactions \cite{Friedman:1991ip, Drell:1970wh}, DFs can be extracted from the data acquired.

Since the first empirical DIS studies, a prodigious number of experiments has been completed and analysed with this purpose \cite{Holt:2010vj, Gao:2017yyd, Geesaman:2018ixo, Ethier:2020way}.
The current status may be described as follows.
(\emph{i}) The wealth of data has enabled development of numerous QCD-related proton DF fits, which are in fair agreement amongst themselves on the domains favoured by dense data.  On the complementary domains, the fits can disagree markedly, with significant physics impact, such as uncertainty in the value of the large-$x$ exponent on the proton's valence-quark distributions \cite{Ball:2016spl, Segarra:2019gbp, Courtoy:2020fex, Abrams:2021xum}.
(\emph{ii}) With phenomenology thus positioned, there is great need for rigorous QCD-connected theory input; but even after roughly fifty years of QCD, this is lacking. Many models have been used to compute valence-quark DFs, but similarities and disagreements speak primarily about the practitioners' assumptions; few predictions are available from realistic solutions of the continuum three-valence-body bound-state problem in QCD \cite{Roberts:2013mja, Chen:2020ijn}; results from lattice-regularised QCD (lQCD) are not yet of sufficient precision to materially contribute to the improvement of DF fitting \cite{Lin:2017snn}; and there are no calculations of the proton's glue distributions.
Plainly, even in the optimal case supplied by the proton, one cannot claim understanding of its DFs.

Whereas the proton is defined by its valence $u+u+d$ quark content, mesons are a different form of hadron matter, being constituted from a valence-quark and \mbox{-a}ntiquark.
%
%
The pion has been a known hadron for more than seventy years \cite{Yukawa:1935xg, Lattes:1947mw}.
Like all mesons, pions are unstable; and while the charged states, $\pi^+ = u\bar d$ ($u$-quark $+$ $d$-antiquark) and $\pi^- = d\bar u$, decay only via weak interactions, they do not live long enough to serve readily as targets.  Hence, regarding parton DFs within the pion, the empirical situation is dire.

Information on the pion's valence-quark DF was obtained in a series of pion-beam experiments at CERN \cite{Corden:1980xf, Badier:1983mj, Betev:1985pg} and Fermilab (E615) \cite{Conway:1989fs} more than thirty years ago and also using the Sullivan process \cite{Sullivan:1971kd} (scattering from the proton's pion cloud) at DESY twenty years ago \cite{Derrick:1996ax, Adloff:1998yg}.  Phenomenological fits of these data \cite{Aicher:2010cb, Sutton:1991ay, Gluck:1999xe, Barry:2018ort, Novikov:2020snp} yield DFs with notable mutual dissimilarities \cite{Chang:2020rdy} and, excepting Ref.\,\cite{Aicher:2010cb}, possessing large-$x$ behaviour in conflict with predictions derived from QCD.  Specifically \cite[Sec.\,5A]{Roberts:2021nhw}: at any resolving scale, $\zeta$, for which data may be interpreted in terms of DFs, the pion's valence-quark DF is predicted to exhibit the following behaviour:
\begin{equation}
\label{pionPDFlargex}
{\mathpzc q}^{\pi}(x;\zeta) \stackrel{x\simeq 1}{=} {\mathpzc c}(\zeta)\, (1-x)^{\beta_\pi(\zeta)}\,,\; \beta_\pi(\zeta)>2\,,
\end{equation}
${\mathpzc c}(\zeta)$ is a constant, \emph{i.e}.\ independent of $x$.  Contradicting this, the $x\simeq 1$ behaviour of the phenomenological DFs in Refs.\,\cite{Sutton:1991ay, Gluck:1999xe, Barry:2018ort, Novikov:2020snp} corresponds to $\beta_\pi(\zeta) \approx 1$.

A key difference between the study of proton and pion DFs is that QCD theory has made real progress with the pion during the past decade.  Preceding this period, numerous models had been used to compute the pion's valence-quark DF \cite{Holt:2010vj}, with the outcome depending on the model chosen.  Today, a unified understanding has been provided for those diverse results \cite{Roberts:2021nhw};
new models are being explored \cite{deTeramond:2018ecg, Lan:2019rba, Ma:2019agv, Chang:2020kjj, Han:2020vjp};
and continuum and lattice studies are delivering parameter-free predictions, with results even for the pion's glue distribution \cite{Ding:2019lwe, Cui:2020tdf, Fan:2021bcr}.  Complementing
these theory developments, new-generation facilities
are developing the capacity to test the predictions via the Sullivan process \cite{Sullivan:1971kd} in high-luminosity electron+proton collisions \cite{JlabTDIS1, JlabTDIS2, Aguilar:2019teb, Chen:2020ijn, Arrington:2021biu} or directly with high-energy, high-intensity pion and kaon beams \cite{Andrieux:2020}.

\smallskip

\noindent\emph{2.$\;$Hadron Scale Pion Distributions}.\,---\,%
In the Standard Model, the pion is a bound state, just like the proton.  The only mechanical difference is that the pion emerges as a pole in the quark+antiquark scattering matrix whereas the proton appears in the study of three-quark scattering.  This perspective defines both modern continuum and lattice treatments, \emph{e.g}.\ Refs.\,\cite{Maris:1997hd, Noaki:2007es}.  It replaces earlier model notions that linked the pion with a long-wavelength fluctuation of a chiral condensate \cite{Brodsky:2012ku}.

Pion properties can be studied using continuum Schwinger function methods (CSMs) \cite{Horn:2016rip, Eichmann:2016yit, Burkert:2017djo, Qin:2020rad}.  Here, QCD's gap and Bethe-Salpeter equations are central.  They have been used to elucidate connections between the pion's Bethe-Salpeter amplitude and light-front wave function \cite{Chang:2013pq}, leading to the prediction that at the hadron scale, $\zeta_H$, with accuracy that will not be exceeded in foreseeable experiments, the pion's leading-twist two dressed-particle distribution amplitude (DA), $\varphi_\pi$, and valence-quark DF are related as follows \cite[Sec.\,3A]{Roberts:2021nhw}:
\begin{align}
\label{DFDA2}
{\mathpzc q}^\pi(x;\zeta_H) & \approx \varphi_\pi^2(x;\zeta_H)/{\mathpzc n}_\varphi\,, 
%
\end{align}
${\mathpzc n}_\varphi = \int_0^1dx\, \varphi_\pi^2(x;\zeta_H)$.
Consequently, $\zeta_H$ is the scale at which the dressed-quark and -antiquark, produced by QCD's gap equation, carry all properties of the pion, including baryon number and momentum:
\begin{equation}
\int_0^1 dx \, {\mathpzc q}^\pi(x;\zeta_H) = 1 = \int_0^1 dx \,2x\, {\mathpzc q}^\pi(x;\zeta_H)\,.
\end{equation}
The pion's sea and glue distributions are zero at $\zeta_H$.  (In the ${\mathpzc G}$-parity symmetry-limit \cite{Lee:1956sw}, a good approximation in Nature, the $\pi^+$ is characterised by a single valence distribution: ${\mathpzc q}^\pi(x;\zeta_H) = {\mathpzc u}^\pi(x;\zeta_H) =\bar{\mathpzc d}^\pi(x;\zeta_H)$.)

Eq.\,\eqref{DFDA2} is important because, following thirty years of controversy, the past decade has seen a clear picture of the pion DA emerge \cite{Chang:2013pq, Stefanis:2015qha, Raya:2015gva, Gao:2016jka, Ding:2019lwe, Cui:2020tdf, Zhong:2021epq, Zhang:2017bzy, Chen:2017gck, Bali:2019dqc}:
$\varphi_\pi$ is a concave function, which is simultaneously dilated and reduced in maximum magnitude relative to its asymptotic profile $\varphi_{\rm as}(x)=6 x(1-x)$ \cite{Lepage:1979zb, Efremov:1979qk, Lepage:1980fj}.  Both features owe to the phenomenon of emergent hadron mass \cite{Roberts:2021xnz}, the likely explanation for $>98$\% of the visible mass in the Universe.

Against this canvas, informed by gap and Bethe-Salpeter equation solutions obtained with refined kernels \cite{Chang:2011ei, Binosi:2016wcx}, Ref.\,\cite{Cui:2020tdf} presented a parameter-free prediction for the pion's valence-quark DF:
\begin{align}
{\mathpzc q}^\pi& (x;\zeta_H)  =
375.32 x^2 (1-x)^2 \nonumber \\
& \times \left[ 1 - 2.5088  \sqrt{x(1-x)} + 2.0250 x (1-x) \right]^2\,.
\end{align}

Importantly, and distinct from modelling and data fitting, the hadron (initial) scale is not introduced as a parameter in Ref.\,\cite{Cui:2020tdf}.  Instead, the value of $\zeta_H$ is a prediction derived from the behaviour of QCD's process-independent (PI) effective charge \cite{Binosi:2016nme, Cui:2019dwv}. 

QCD's PI charge is accurately interpolated by the following function \cite{Cui:2020tdf}:
\begin{align}
\label{Eqhatalpha}
\hat{\alpha}(k^2) & = \frac{\gamma_m \pi}{\ln\left[\frac{{\mathpzc K}^2(k^2)}{\Lambda_{\rm QCD}^2}\right]}
\,,\; {\mathpzc K}^2(y) = \frac{a_0^2 + a_1 y + y^2}{b_0 + y}\,,
\end{align}
$\gamma_m=4/[11 - (2/3)n_f]$, $n_f=4$, $\Lambda_{\rm QCD}=0.234\,$GeV, with (in GeV$^2$): $a_0=0.104(1)$, $a_1=0.0975$, $b_0=0.121(1)$.  Here, to simplify comparisons with other sources, the interpolation is expressed through $\Lambda_{\rm QCD}$, the mass-scale introduced in perturbation theory via the process of regularisation and renormalisation required in defining four-dimensional quantum field theories.

$\hat\alpha(k^2)$ agrees to better than 0.1\% with perturbative QCD's one-loop coupling on $k^2 \gtrsim (9\Lambda_{\rm QCD})^2$.  That perturbative coupling exhibits an unphysical (Landau) pole at $k^2=\Lambda_{\rm QCD}^2$ \cite{Deur:2016tte}; but in deriving $\hat\alpha(k^2)$, the Landau pole is seen to be eliminated by nonperturbative gauge-sector dynamics \cite{Binosi:2016nme, Cui:2019dwv}.
Comparing with the perturbative expression, $k^2/\Lambda_{\rm QCD}^2 \to {\mathpzc K}^2(k^2)/\Lambda_{\rm QCD}^2$ as the logarithm's argument in Eq.\,\eqref{Eqhatalpha}.  The value
$m_G := {\mathpzc K}(k^2=\Lambda_{\rm QCD}^2) = 0.331(2)\,{\rm GeV}$
defines a screening mass.  It marks a boundary: the coupling alters character at $k \simeq m_G$ so that modes with $k^2 \lesssim m_G^2$ are screened from interactions and the theory enters a practically conformal domain \cite{Brodsky:2008be, Aguilar:2015bud, Gao:2017uox, Huber:2018ned}.  The line $k=m_G$ separates long- and short-wavelength physics; hence, serves as the natural definition for the hadron scale, \emph{viz}.\ $\zeta_H=m_G$.
%

\smallskip

\noindent\emph{3.$\;$Pion Distributions at $\zeta=2\,$GeV}.\,---\,%
%
DFs and/or their moments are commonly quoted at $\zeta=\zeta_2=2\,$GeV; and one may obtain the DFs at $\zeta_2$ via evolution from another scale by integrating the DGLAP equations \cite{Dokshitzer:1977sg, Gribov:1972ri, Lipatov:1974qm, Altarelli:1977zs}.
When completing this exercise, a prescription must be specified because the standard DGLAP equations involve QCD's running coupling.  In modelling and fitting, it is usual to adopt a purely perturbative-QCD perspective, implementing evolution with a DGLAP kernel calculated at a given order in perturbation theory.  If the scale at which evolution begins is large enough, then leading-order (LO) evolution kernels may be adequate, at least in practice.  If they fail, then next-to-leading-order (NLO) can be implemented, and so on, in principle.

\begin{figure}[t]
\includegraphics[width=0.44\textwidth]{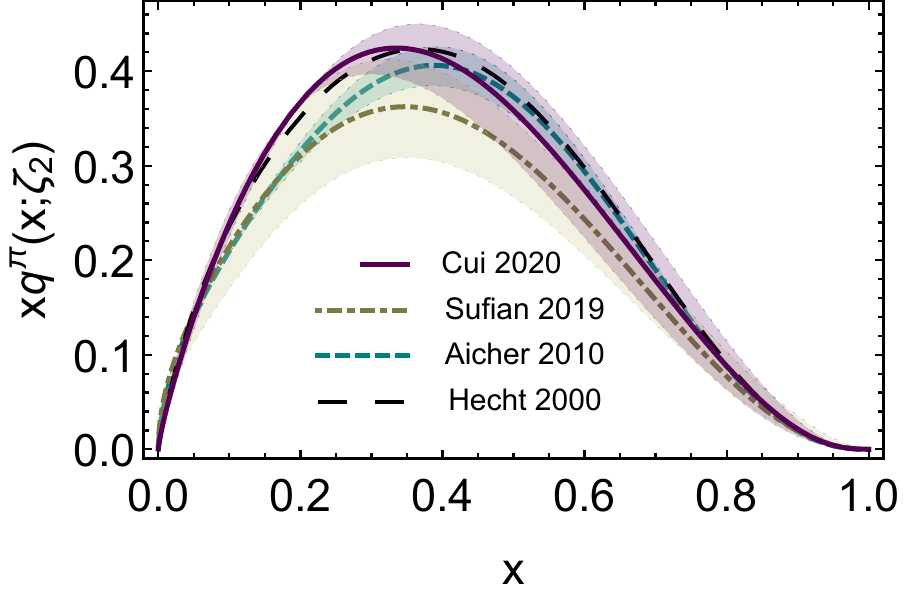}
\caption{\label{FvalenceA}
$x {\mathpzc q}^\pi(x,\zeta_2=2\,{\rm GeV})$, pion valence-quark distribution: solid purple curve, prediction from Ref.\,\cite{Cui:2020tdf}.
The associated band expresses a conservative estimate of uncertainty in the prediction, obtained by varying $\zeta_H$ by $\pm 10$\%.
Comparisons:
dot-dashed olive-green curve, lQCD result \cite{Sufian:2019bol};
short-dashed teal curve and associated uncertainty band, fit to E615 data \cite{Aicher:2010cb}, which incorporated next-to-leading-logarithm effects in the hard-scattering kernel;
and long-dashed black curve, original CSM result \cite{Hecht:2000xa}.
}
\end{figure}

\begin{figure}[t]
\vspace*{2ex}

\leftline{\hspace*{0.5em}{\large{\textsf{A}}}}
\vspace*{-4ex}
\includegraphics[width=0.44\textwidth]{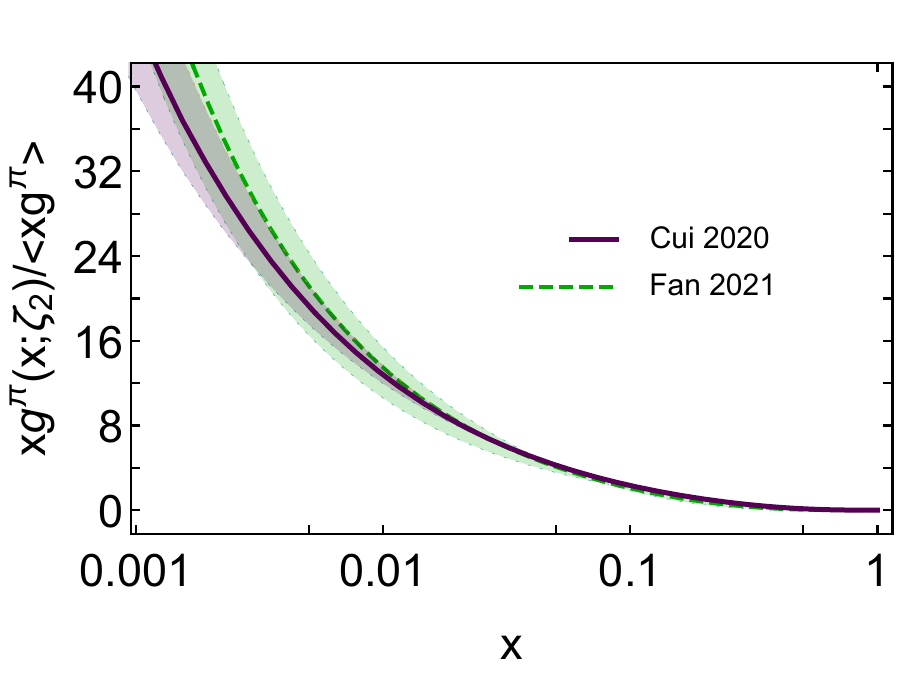}
\vspace*{-1ex}

\leftline{\hspace*{0.5em}{\large{\textsf{B}}}}
\vspace*{-4ex}
\includegraphics[width=0.44\textwidth]{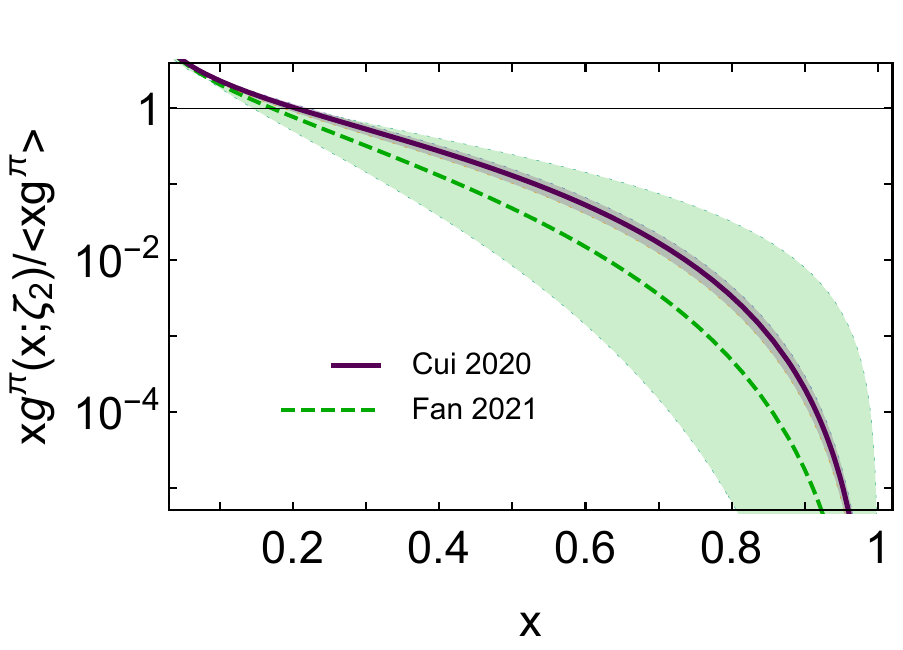}

\caption{\label{FlQCDglue}
$x {\mathpzc g}^\pi(x,\zeta_2)/\langle x{\mathpzc g}^\pi_{\zeta_2}\rangle$, $\zeta_2=2\,$GeV, normalised glue distribution: solid purple curve, prediction from Ref.\,\cite{Cui:2020tdf}.
Panel A highlights low-$x$ and Panel B, large-$x$.
Comparison curve: lQCD result \cite{Fan:2021bcr}, for which the associated band expresses aspects of statistical error and systematic uncertainty associated with using an assumed functional form to fit the results.
}
\end{figure}

A different approach is advocated in Refs.\,\cite{Cui:2019dwv, Cui:2020tdf}.  Namely, adapting ideas from Refs.\,\cite{Grunberg:1982fw, Dokshitzer:1998nz, Deur:2016tte}, evolution is implemented by using $\hat\alpha(k^2)$  in Eq.\,\eqref{Eqhatalpha} to integrate the one-loop DGLAP equations.  This $\hat\alpha$ scheme eliminates ambiguity from the resulting predictions because it renders moot any questions regarding the order of the evolution kernels.  Working with the hadron scale DFs described in Sec.\,2, one obtains the parameter-free predictions for the  $\zeta_2=2\,{\rm GeV}$ valence, sea, and glue DFs drawn in Figs.\,\ref{FvalenceA}\,--\,\ref{FlQCDglue}.  Using those predictions, one calculates the following light-front momentum fractions:
\begin{align}
\label{momfracs}
&\begin{array}{c|c|c}
\langle 2 x {\mathpzc q}^\pi_{\zeta_2}\rangle & \langle x {\mathpzc S}^\pi_{\zeta_2}\rangle &
\langle x {\mathpzc g}^\pi_{\zeta_2}\rangle\\\hline
0.48(4) & 0.11(2) & 0.41(2)
\end{array}\,,\\
\mbox{where}\; &  \langle x^n {\mathpzc p}^\pi_{\zeta_2}\rangle = \int_0^1 dx\, x^n {\mathpzc p}^\pi(x;\zeta_2).
\end{align}
%
Evidently, gluon partons carry a large fraction of the pion's momentum at $\zeta_2$.

\begin{figure}[t]
\includegraphics[width=0.44\textwidth]{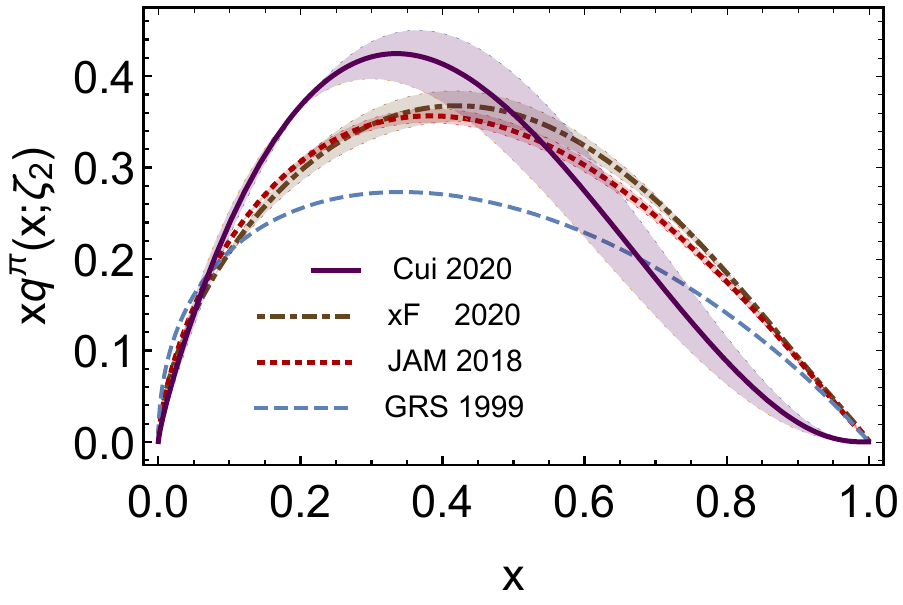}
\caption{\label{Fvalence}
$x {\mathpzc q}^\pi(x,\zeta_2=2\,{\rm GeV})$, pion valence-quark distribution: solid purple curve and associated uncertainty band, prediction from Ref.\,\cite{Cui:2020tdf}.
%
%
Comparisons are selected fits to data: dashed blue curve, \cite{Gluck:1999xe}; dotted red curve and associated uncertainty band, \cite{Barry:2018ort}; dot-dashed brown curve and band, \cite{Novikov:2020snp}.
}
\end{figure}

Given that the calculation of Mellin moments of the pion's valence-quark DF is still the subject of many lQCD studies \cite{Joo:2019bzr, Gao:2020ito}, it is worth recording the following predictions obtained using the pion DFs in Ref.\,\cite{Cui:2020tdf}:
\begin{subequations}
\label{momfracsV}
\begin{align}
\begin{array}{c|c|c}
\langle x {\mathpzc q}_{\zeta_2}^\pi\rangle & \langle x^2 {\mathpzc q}^\pi_{\zeta_2}\rangle
& \langle x^3 {\mathpzc q}^\pi_{\zeta_2}\rangle \\\hline
0.24(2) & 0.094(13) & 0.047(08)
\end{array}\,,\\
\begin{array}{c|c|c}
\langle x^4 {\mathpzc q}^\pi_{\zeta_2}\rangle &
\langle x^5 {\mathpzc q}^\pi_{\zeta_2}\rangle& \langle x^6 {\mathpzc q}^\pi_{\zeta_2}\rangle
 \\\hline
0.027(05) & 0.017(04) & 0.011(03)
\end{array}\,.
\end{align}
\end{subequations}
Larger values for the $n\geq 2$ moments indicate a harder DF, \emph{i.e}.\ a function with greater support on the valence-quark domain than the prediction in Ref.\,\cite{Cui:2020tdf}; consequently, less support on the complementary domain.

It is worth continuing with a discussion of the pointwise behaviour of the pion's valence-quark distribution, Fig.\,\ref{FvalenceA}.
Within uncertainties, the parameter-free prediction from Ref.\,\cite{Cui:2020tdf} agrees with the lQCD result in Ref.\,\cite{Sufian:2019bol}; and both agree well on the valence-quark domain with the NLO analysis of E615 data \cite{Conway:1989fs} described in Ref.\,\cite{Aicher:2010cb}, which is the only work thus far to include next-to-leading-logarithmic (NLL) threshold resummation effects when calculating the hard-scattering kernel.
(N.B.\, Ref.\,\cite{Aicher:2010cb} used sea and glue distributions from Ref.\,\cite{Gluck:1999xe}; hence, the valence distribution in Ref.\,\cite{Aicher:2010cb} is likely less reliable on $x\lesssim 0.4$.)
Interestingly, the result in Ref.\,\cite{Hecht:2000xa}, which reignited debate over the pion's valence-quark DF and was obtained using algebraic \emph{Ans\"atze} for the elements in the pion wave function, 
also agrees with the modern CSM prediction.
The continuum prediction \cite{Cui:2020tdf}, lQCD result \cite{Sufian:2019bol}, NLO+NLL fit \cite{Aicher:2010cb}, and QCD-based model \cite{Hecht:2000xa} are consistent with Eq.\,\eqref{pionPDFlargex}.  (See also Ref.\,\cite[Fig.\,5a]{Cui:2020tdf}.)
%

\begin{figure}[t]
%
\includegraphics[width=0.44\textwidth]{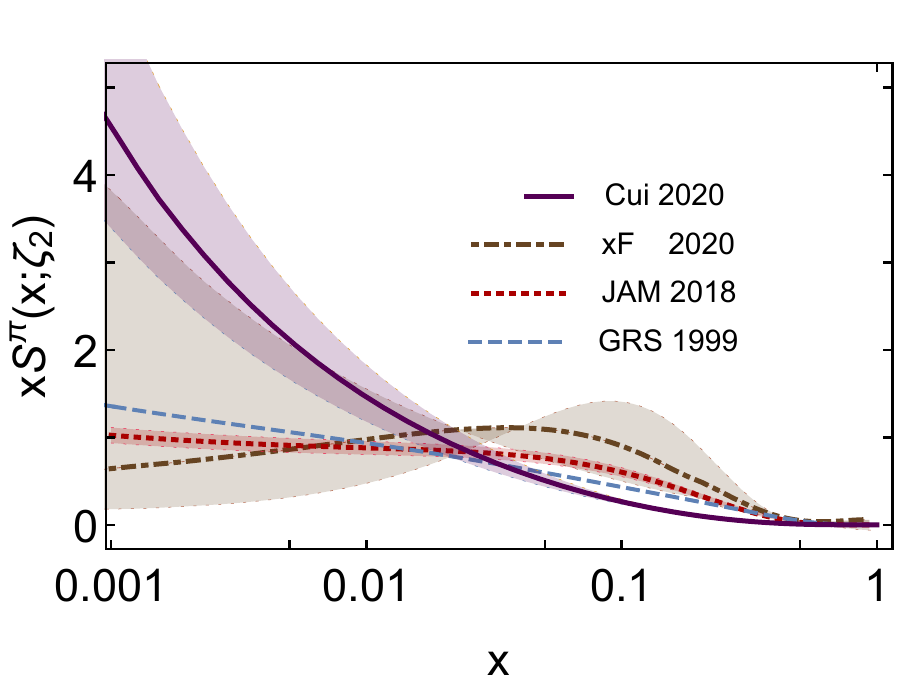}
\caption{\label{Fsea}
Sea distribution, $x {\mathpzc S}^\pi(x,\zeta_2=2\,{\rm GeV})$: solid purple curve, prediction from Ref.\,\cite{Cui:2020tdf}.
The associted band expresses a conservative estimate of uncertainty in the prediction, obtained by varying $\zeta_H$ by $\pm 10$\%.
Comparisons are selected fits to data: dashed blue curve, \cite{Gluck:1999xe}; dotted red curve and associated band, \cite{Barry:2018ort}; dot-dashed brown curve and band, \cite{Novikov:2020snp}.
}
\end{figure}

\begin{figure}[t]
\vspace*{2ex}

\leftline{\hspace*{0.5em}{\large{\textsf{A}}}}
\vspace*{-4ex}
\includegraphics[width=0.44\textwidth]{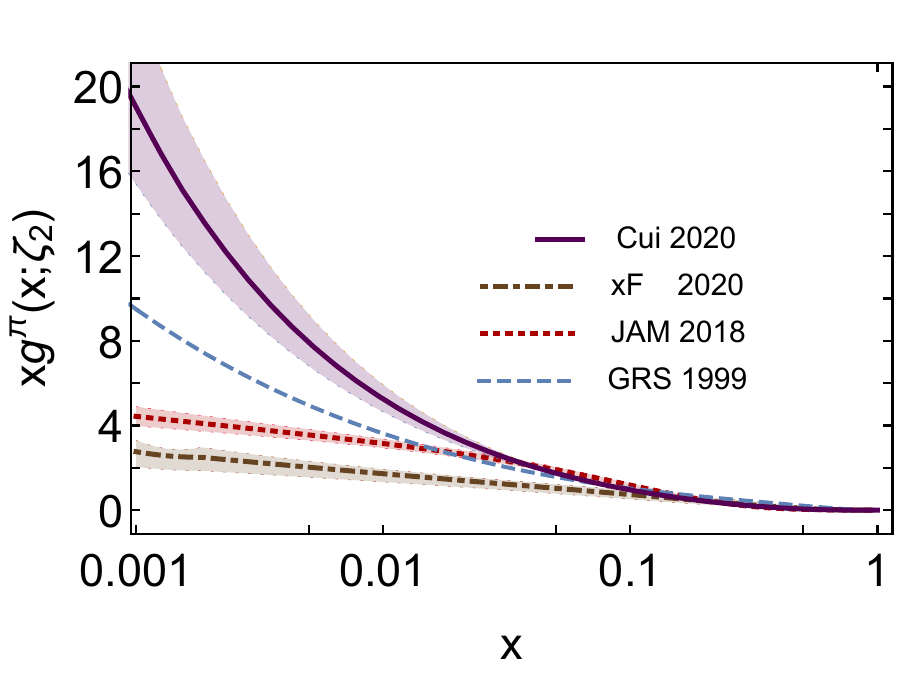}
\vspace*{-1ex}

\leftline{\hspace*{0.5em}{\large{\textsf{B}}}}
\vspace*{-4ex}
\includegraphics[width=0.44\textwidth]{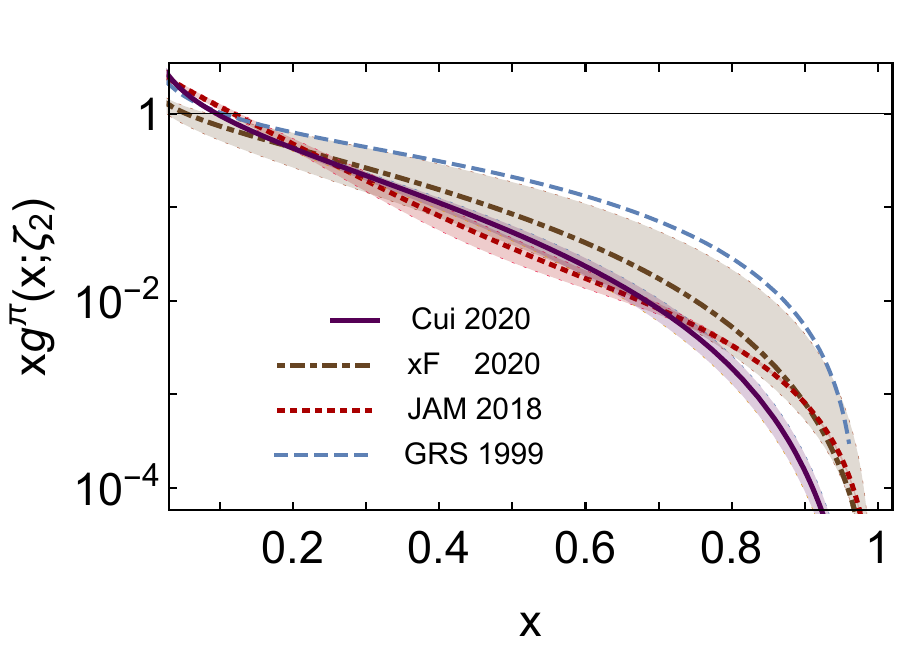}

\caption{\label{Fglue}
Glue distribution, $x {\mathpzc g}^\pi(x,\zeta_2=2\,{\rm GeV})$: solid purple curve, prediction from Ref.\,\cite{Cui:2020tdf}.  Panel A highlights low-$x$ and Panel B, large-$x$.
The band surrounding this curve expresses a conservative estimate of uncertainty in the prediction, obtained by varying $\zeta_H$ by $\pm 10$\%.
Comparisons are selected fits to data: dashed blue curve, \cite{Gluck:1999xe}; dotted red curve and associated band, \cite{Barry:2018ort}; dot-dashed brown curve and band, \cite{Novikov:2020snp}.
}
\end{figure}

On the other hand, NLL effects were omitted in producing the comparison fits \cite{Gluck:1999xe, Barry:2018ort, Novikov:2020snp} drawn explicitly in Fig.\,\ref{Fvalence}, which display some notable mutual differences and are also in conflict with Eq.\,\eqref{pionPDFlargex}.
Since the fits in Refs.\,\cite{Gluck:1999xe, Barry:2018ort, Novikov:2020snp} are harder (possess greater support) at large-$x$ than the prediction from Ref.\,\cite{Cui:2020tdf}, then one can anticipate that they must also produce harder sea and glue distributions because the valence distribution serves as a source for both sea and glue.  Equally, owing to momentum conservation, the data fits must possess less support at low $x$.  These expectations were confirmed in Ref.\,\cite[Fig.\,4b]{Cui:2020tdf} and the results are highlighted by Figs.\,\ref{Fsea}, \ref{Fglue} herein.

Regarding Fig.\,\ref{Fglue}, it is worth emphasising that at $\zeta=\zeta_2$ the pion's predicted glue distribution is larger in magnitude than its valence distribution on $x\lesssim 0.22$.  Such an outcome was to be expected.  Indeed, as stated in connection with Eq.\,\eqref{Eqhatalpha}, at $\zeta=\zeta_H$ all  properties of the pion are expressed in the dressed-quark and -antiquark degrees-of-freedom that are generated dynamically in solving the continuum bound-state problem.  Under evolution to $\zeta>\zeta_H$, these quasiparticles steadily shed their cladding, thereby producing populations of sea and glue partons.  The size of these populations increases with increasing $\zeta$.  Hence, all symmetry-preserving treatments of pion structure and dynamics must produce nonzero valence, sea, and glue DFs at any scale for which an analysis of experiments may be interpreted in terms of DFs.  
Furthermore, there is no scale at which all three vanish identically.

Figs.\,\ref{FvalenceA}\,--\,\ref{Fglue} show that the pointwise form of each DF obtained via a fit to data which omits NLL effects disagrees markedly with the comparable CSM prediction \cite{Cui:2020tdf}.  This stresses the need for both improved data analyses\footnote{Complementing Ref.\,\cite{Aicher:2010cb}, the potential impacts of including NLL effects when extracting DFs from data are also illustrated in Ref.\,\cite{patrick_barry_2020_4019411}, \emph{e.g}.\ all DFs are softer at large-$x$ and a greater fraction of the pion's light-front momentum is carried by glue, with the cost paid largely by the sea fraction.} and additional rigorous DF predictions from QCD theory.
%

In this connection, recalling Fig.\,\ref{FvalenceA}, the CSM prediction for the pion's valence-quark DF \cite{Cui:2020tdf} is confirmed by the lQCD result in Ref.\,\cite{Sufian:2019bol}.

No lQCD results are available for the pion's sea distribution; so, nothing can yet be directly concluded on this score.  Regarding phenomenology, Fig.\,\ref{Fsea} highlights that the pion's sea distribution is very poorly constrained by existing data analyses.

On the other hand, lQCD results for the pion's glue DF have recently become available \cite{Fan:2021bcr}. They are compared in Fig.\,\ref{FlQCDglue} with the CSM prediction \cite{Cui:2020tdf}: within uncertainties, there is pointwise agreement between the two results on the entire depicted domains.
Combined with similar agreement between CSM and lQCD predictions for the pion's valence distribution (Fig.\,\ref{FvalenceA}), one has evidence to suggest that the CSM prediction for the pion's sea distribution, Fig.\,\ref{Fsea}, is also reliable.

Such quantitative agreement between continuum and lattice results for pion DFs supports the perspective on the origin of hadron masses developed in Ref.\,\cite{Roberts:2016vyn}.  Namely, the mass of the pion is much smaller than that of the $\rho$-meson (and the proton) owing to a symmetry-ensured cancellation in this near Nambu-Goldstone boson between the positive one-body-dressing content of the QCD trace anomaly and the negative binding energy produced by interactions.   (See also Ref.\,\cite[Sec.\,2\,D]{Roberts:2021nhw}.)

\smallskip

\noindent\emph{4.$\;$Summary and Perspective}.\,---\,%
Through comparisons with continuum and lattice predictions for pion distribution functions (DFs), evidence is accumulating which indicates that the pion DFs inferred from analyses of existing data are pointwise inaccurate, possessing too much support on the valence-quark domain ($x\gtrsim 0.2$) and too little on its complement.  Additional evidence suggests that the discrepancies may be ameliorated, or even eliminated, by inclusion of next-to-leading-logarithmic threshold resummation effects in the hard-scattering kernels whose accurate representation is an essential element in the determination of DFs through data fitting.  If these remarks are correct, then their implications for the DFs of nucleons and nuclei should similarly be considered; especially because imperfect extractions of such DFs may obscure or provide misleading signals of physics beyond the Standard Model.
In any event, one must look forward to the era in which sound QCD-connected predictions are exploited in providing material constraints on data analyses.
This time is approaching for pions and kaons.

As Nature's most basic Nambu-Goldstone bosons, pions and kaons are unique.  Being mesons -- quark+antiquark bound-states -- they also typify a form of hadron matter about whose structure very little is empirically known.  That is changing, with new-generation facilities, in operation or planning, having the capacity to provide unprecedented access to these systems as \emph{targets}.

Critically, theory is also making rapid progress, beginning to deliver robust predictions for the measurable properties of pions and kaons.  So, it is reasonable to expect that sometime after the centenary of the prediction of the pion's existence \cite{Yukawa:1935xg} and before the centenary of its discovery \cite{Lattes:1947mw}, science will finally have the information necessary to draw charts which reveal the pion's structure in exquisite detail.  This may finally explain why, \emph{inter alia}, basic features of emergent hadron mass are hidden in the pion whereas they are blatant in the proton.


\smallskip

%
\noindent\emph{Acknowledgments}.\,---\,%
We are grateful to Z.~Fan and H.-W.~Lin for providing us with information on their simulation results;
to J.~Lan and Z.-B.~Xing for assistance with running DGLAP evolution codes;
and for constructive comments from
V.~Andrieux, D.~Binosi, W.-C.~Chang, Z.-F.~Cui, O.~Denisov, M.~Ding, R.~Ent, F.~Gao, T.~Horn, W.-D.~Nowak, S.~Platchkov, C.~Quintans, K.~Raya, J.~Rodr{\'{\i}}guez-Quintero and S.\,M.~Schmidt.
This work benefited from discussions and presentations during the ``Workshop on Pion and Kaon Structure at the EIC'', hosted by the Center for Frontiers in Nuclear Science, 2020 Jun.\ 2-5, and the ongoing series of CERN-hosted workshops on ``Perceiving the Emergence of Hadron Mass through AMBER@CERN'' -- 2019 Dec.\ 11 (I), 2020 Mar.\ 30-Apr.\ 02 (II), 2020 Aug.\ 06-07 (III), 2020 Nov.\ 30-Dec.\ 4 (IV), 2021 Apr.\ 27-30 (V).
Research supported by:
Government of China \emph{Thousand Talents Plan for Young Professionals}.
%


\begin{thebibliography}{10}
\expandafter\ifx\csname url\endcsname\relax
  \def\url#1{\texttt{#1}}\fi
\expandafter\ifx\csname urlprefix\endcsname\relax\def\urlprefix{URL }\fi
\expandafter\ifx\csname href\endcsname\relax
  \def\href#1#2{#2} \def\path#1{#1}\fi

\bibitem{RutherfordIV}
E.~Rutherford, Collision of alpha particles with light atoms iv. an anomalous
  effect in nitrogen, Philosophical Magazine xxxvii (1919) 581--587.

\bibitem{Cockroft:1932I}
J.~D. Cockroft, E.~T.~S. Walton, {Experiments with High Velocity Positive Ions.
  I. Further Developments in the Method of Obtaining High Velocity Positive
  Ions}, Proc. Roy. Soc. Lond. A 136 (1932) 619--630.

\bibitem{Taylor:1991ew}
R.~E. Taylor, {Deep inelastic scattering: The Early years}, Rev. Mod. Phys. 63
  (1991) 573--595.

\bibitem{Kendall:1991np}
H.~W. Kendall, {Deep inelastic scattering: Experiments on the proton and the
  observation of scaling}, Rev. Mod. Phys. 63 (1991) 597--614.

\bibitem{Friedman:1991nq}
J.~I. Friedman, {Deep inelastic scattering: Comparisons with the quark model},
  Rev. Mod. Phys. 63 (1991) 615--629.

\bibitem{Marciano:1979wa}
W.~J. Marciano, H.~Pagels, {Quantum Chromodynamics}, Nature 279 (1979)
  479--483.

\bibitem{Brock:1993sz}
R.~Brock, et~al., {Handbook of perturbative QCD: Version 1.0}, Rev. Mod. Phys.
  67 (1995) 157--248.

\bibitem{Friedman:1991ip}
J.~I. Friedman, H.~W. Kendall, R.~E. Taylor, {Deep inelastic scattering:
  Acknowledgements}, Rev. Mod. Phys. 63 (1991) 629.

\bibitem{Drell:1970wh}
S.~Drell, T.-M. Yan, {Massive Lepton Pair Production in Hadron-Hadron
  Collisions at High-Energies}, Phys. Rev. Lett. 25 (1970) 316--320, [Erratum:
  Phys. Rev. Lett. \textbf{25}, 902 (1970)].

\bibitem{Holt:2010vj}
R.~J. Holt, C.~D. Roberts, {Distribution Functions of the Nucleon and Pion in
  the Valence Region}, Rev. Mod. Phys. 82 (2010) 2991--3044.

\bibitem{Gao:2017yyd}
J.~Gao, L.~Harland-Lang, J.~Rojo, {The Structure of the Proton in the LHC
  Precision Era}, Phys. Rept. 742 (2018) 1--121.

\bibitem{Geesaman:2018ixo}
D.~F. Geesaman, P.~E. Reimer, {The sea of quarks and antiquarks in the
  nucleon}, Rept. Prog. Phys. 82 (2019) 046301.

\bibitem{Ethier:2020way}
J.~J. Ethier, E.~R. Nocera, {Parton Distributions in Nucleons and Nuclei}, Ann.
  Rev. Nucl. Part. Sci. 70 (2020) 43--76.

\bibitem{Ball:2016spl}
R.~D. Ball, E.~R. Nocera, J.~Rojo, {The asymptotic behaviour of parton
  distributions at small and large $x$}, Eur. Phys. J. C 76 (2016) 383.

\bibitem{Segarra:2019gbp}
E.~Segarra, A.~Schmidt, T.~Kutz, D.~Higinbotham, E.~Piasetzky, M.~Strikman,
  L.~Weinstein, O.~Hen, {Neutron Valence Structure from Nuclear Deep Inelastic
  Scattering}, Phys. Rev. Lett. 124 (2020) 092002.

\bibitem{Courtoy:2020fex}
A.~Courtoy, P.~M. Nadolsky, {Testing momentum dependence of the nonperturbative
  hadron structure in a global QCD analysis}, Phys. Rev. D 103 (2021) 054029.

\bibitem{Abrams:2021xum}
D.~Abrams, et~al., {Measurement of the Nucleon $F^n_2/F^p_2$ Structure Function
  Ratio by the Jefferson Lab MARATHON Tritium/Helium-3 Deep Inelastic
  Scattering Experiment -- arXiv:2104.05850 [hep-ex]}.

\bibitem{Roberts:2013mja}
C.~D. Roberts, R.~J. Holt, S.~M. Schmidt, {Nucleon spin structure at very high
  $x$}, Phys. Lett. B 727 (2013) 249--254.

\bibitem{Chen:2020ijn}
X.~Chen, F.-K. Guo, C.~D. Roberts, R.~Wang, {Selected Science Opportunities for
  the EicC}, Few Body Syst. 61 (2020) 43.

\bibitem{Lin:2017snn}
H.-W. Lin, et~al., {Parton distributions and lattice QCD calculations: a
  community white paper}, Prog. Part. Nucl. Phys. 100 (2018) 107--160.

\bibitem{Yukawa:1935xg}
H.~Yukawa, {On the interaction of elementary particles}, Proc. Phys. Math. Soc.
  Jap. 17 (1935) 48--57.

\bibitem{Lattes:1947mw}
C.~M.~G. Lattes, H.~Muirhead, G.~P.~S. Occhialini, C.~F. Powell, {Processes
  involving charged mesons}, Nature 159 (1947) 694--697.

\bibitem{Corden:1980xf}
M.~Corden, et~al., {Production of Muon Pairs in the Continuum Region by
  39.5-{GeV}/$c \pi^\pm$, $K^\pm$, $p$ and $\bar{p}$ Beams Incident on a
  Tungsten Target}, Phys. Lett. B 96 (1980) 417--421.

\bibitem{Badier:1983mj}
J.~Badier, et~al., {Experimental determination of the {$\pi$}-meson structure
  functions by the Drell-Yan mechanism}, Z. Phys. C 18 (1983) 281.

\bibitem{Betev:1985pg}
B.~Betev, et~al., {Observation of anomalous scaling violation in muon pair
  production by 194-GeV/c {$\pi$}-tungsten interactions}, Z. Phys. C 28 (1985)
  15.

\bibitem{Conway:1989fs}
J.~S. Conway, et~al., {Experimental study of muon pairs produced by 252-GeV
  pions on tungsten}, Phys. Rev. D 39 (1989) 92--122.

\bibitem{Sullivan:1971kd}
J.~D. Sullivan, {One pion exchange and deep inelastic electron - nucleon
  scattering}, Phys. Rev. D 5 (1972) 1732--1737.

\bibitem{Derrick:1996ax}
M.~Derrick, et~al., {Observation of events with an energetic forward neutron in
  deep inelastic scattering at HERA}, Phys. Lett. B 384 (1996) 388--400.

\bibitem{Adloff:1998yg}
C.~Adloff, et~al., {Measurement of leading proton and neutron production in
  deep inelastic scattering at HERA}, Eur. Phys. J. C 6 (1999) 587--602.

\bibitem{Aicher:2010cb}
M.~Aicher, A.~Sch{\"a}fer, W.~Vogelsang, {Soft-Gluon Resummation and the
  Valence Parton Distribution Function of the Pion}, Phys.\ Rev.\ Lett. 105
  (2010) 252003.

\bibitem{Sutton:1991ay}
P.~J. Sutton, A.~D. Martin, R.~G. Roberts, W.~J. Stirling, {Parton
  distributions for the pion extracted from Drell-Yan and prompt photon
  experiments}, Phys. Rev. D 45 (1992) 2349--2359.

\bibitem{Gluck:1999xe}
M.~Gl{\"u}ck, E.~Reya, I.~Schienbein, {Pionic parton distributions revisited},
  Eur. Phys. J. C 10 (1999) 313--317.

\bibitem{Barry:2018ort}
P.~C. Barry, N.~Sato, W.~Melnitchouk, C.-R. Ji, {First Monte Carlo Global QCD
  Analysis of Pion Parton Distributions}, Phys. Rev. Lett. 121 (2018) 152001.

\bibitem{Novikov:2020snp}
I.~Novikov, et~al., {Parton Distribution Functions of the Charged Pion Within
  The xFitter Framework}, Phys. Rev. D 102 (2020) 014040.

\bibitem{Chang:2020rdy}
W.-C. Chang, J.-C. Peng, S.~Platchkov, T.~Sawada, {Constraining gluon density
  of pions at large $x$ by pion-induced $J/\psi$ production}, Phys. Rev. D 102
  (2020) 054024.

\bibitem{Roberts:2021nhw}
C.~D. Roberts, D.~G. Richards, T.~Horn, L.~Chang, {\emph{Insights into the
  Emergence of Mass from Studies of Pion and Kaon Structure} --
  arXiv:2102.01765 [hep-ph]}, Prog. Part. Nucl. Phys. \emph{in press}.

\bibitem{deTeramond:2018ecg}
G.~F. de~Teramond, T.~Liu, R.~S. Sufian, H.~G. Dosch, S.~J. Brodsky, A.~Deur,
  {Universality of Generalized Parton Distributions in Light-Front Holographic
  QCD}, Phys. Rev. Lett. 120 (2018) 182001.

\bibitem{Lan:2019rba}
J.~Lan, C.~Mondal, S.~Jia, X.~Zhao, J.~P. Vary, {Pion and kaon parton
  distribution functions from basis light front quantization and QCD
  evolution}, Phys. Rev. D 101 (2020) 034024.

\bibitem{Ma:2019agv}
Z.-L. Ma, J.-Q. Zhu, Z.~Lu, {Quasiparton distribution function and
  quasigeneralized parton distribution of the pion in a spectator model}, Phys.
  Rev. D 101 (2020) 114005.

\bibitem{Chang:2020kjj}
L.~Chang, K.~Raya, X.~Wang, {Pion Parton Distribution Function in Light-Front
  Holographic QCD}, Chin. Phys. C 44~(11) (2020) 114105.

\bibitem{Han:2020vjp}
C.~Han, G.~Xie, R.~Wang, X.~Chen, {An Analysis of Parton Distribution Functions
  of the Pion and the Kaon with the Maximum Entropy Input}, Eur. Phys. J. C 81
  (2021) 302.

\bibitem{Ding:2019lwe}
M.~Ding, K.~Raya, D.~Binosi, L.~Chang, C.~D. Roberts, S.~M. Schmidt, {Symmetry,
  symmetry breaking, and pion parton distributions}, Phys. Rev. D 101~(5)
  (2020) 054014.

\bibitem{Cui:2020tdf}
Z.-F. Cui, M.~Ding, F.~Gao, K.~Raya, D.~Binosi, L.~Chang, C.~D. Roberts,
  J.~Rodr\'{\i}guez-Quintero, S.~M. Schmidt, {Kaon and pion parton
  distributions}, Eur. Phys. J. C 80 (2020) 1064.

\bibitem{Fan:2021bcr}
Z.~Fan, H.-W. Lin, {Gluon Parton Distribution of the Pion from Lattice QCD --
  arXiv:2104.06372 [hep-lat]}.

\bibitem{JlabTDIS1}
{C. Keppel, B. Wojtsekhowski, P. King, D. Dutta, J. Annand, J. Zhang \emph{et
  al}.}, Measurement of tagged deep inelastic scattering (\mbox{TDIS})\mbox{
  }approved Jefferson Lab experiment E12-15-006.

\bibitem{JlabTDIS2}
{K. Park, R. Montgomery, T. Horn \emph{et al}.}, Measurement of kaon structure
  function through tagged deep inelastic scattering (\mbox{TDIS})\mbox{
  }approved Jefferson Lab experiment C12-15-006A.

\bibitem{Aguilar:2019teb}
A.~C. Aguilar, et~al., {Pion and Kaon Structure at the Electron-Ion Collider},
  Eur. Phys. J. A 55 (2019) 190.

\bibitem{Arrington:2021biu}
J.~Arrington, et~al., {Revealing the structure of light pseudoscalar mesons at
  the electron\textendash{}ion collider}, J. Phys. G 48 (2021) 075106.

\bibitem{Andrieux:2020}
V.~Andrieux, B.~Parsamyan, {From COMPASS to AMBER: exploring fundamental
  properties of hadrons}, CERN EP Newsletter 2020/12 - 2021/02.

\bibitem{Maris:1997hd}
P.~Maris, C.~D. Roberts, P.~C. Tandy, {Pion mass and decay constant}, Phys.
  Lett. B 420 (1998) 267--273.

\bibitem{Noaki:2007es}
J.~Noaki, S.~Aoki, H.~Fukaya, S.~Hashimoto, T.~Kaneko, H.~Matsufuru, T.~Onogi,
  E.~Shintani, N.~Yamada, {Light meson spectrum with N(f) = 2 dynamical overlap
  fermions}, PoS LATTICE2007 (2007) 126.

\bibitem{Brodsky:2012ku}
S.~J. Brodsky, C.~D. Roberts, R.~Shrock, P.~C. Tandy, {Confinement contains
  condensates}, Phys. Rev. C 85 (2012) 065202.

\bibitem{Horn:2016rip}
T.~Horn, C.~D. Roberts, {The pion: an enigma within the Standard Model}, J.
  Phys. G. 43 (2016) 073001.

\bibitem{Eichmann:2016yit}
G.~Eichmann, H.~Sanchis-Alepuz, R.~Williams, R.~Alkofer, C.~S. Fischer,
  {Baryons as relativistic three-quark bound states}, Prog. Part. Nucl. Phys.
  91 (2016) 1--100.

\bibitem{Burkert:2017djo}
V.~D. Burkert, C.~D. Roberts, {Roper resonance: Toward a solution to the
  fifty-year puzzle}, Rev. Mod. Phys. 91 (2019) 011003.

\bibitem{Qin:2020rad}
S.-X. Qin, C.~D. Roberts, {Impressions of the Continuum Bound State Problem in
  QCD}, Chin. Phys. Lett. 37~(12) (2020) 121201.

\bibitem{Chang:2013pq}
L.~Chang, I.~C. Cloet, J.~J. Cobos-Martinez, C.~D. Roberts, S.~M. Schmidt,
  P.~C. Tandy, {Imaging dynamical chiral symmetry breaking: pion wave function
  on the light front}, Phys. Rev. Lett. 110 (2013) 132001.

\bibitem{Lee:1956sw}
T.~D. Lee, C.-N. Yang, {Charge Conjugation, a New Quantum Number $G$, and
  Selection Rules Concerning a Nucleon Anti-nucleon System}, Nuovo Cim. 10
  (1956) 749--753.

\bibitem{Stefanis:2015qha}
N.~G. Stefanis, A.~V. Pimikov, {Chimera distribution amplitudes for the pion
  and the longitudinally polarized \mbox{$\rho$}-meson}, Nucl. Phys. A 945
  (2016) 248--268.

\bibitem{Raya:2015gva}
K.~Raya, L.~Chang, A.~Bashir, J.~J. Cobos-Martinez, L.~X.
  Guti{\'e}rrez-Guerrero, C.~D. Roberts, P.~C. Tandy, {Structure of the neutral
  pion and its electromagnetic transition form factor}, Phys. Rev. D 93 (2016)
  074017.

\bibitem{Gao:2016jka}
F.~Gao, L.~Chang, Y.-X. Liu, {Bayesian extraction of the parton distribution
  amplitude from the Bethe-Salpeter wave function}, Phys. Lett. B 770 (2017)
  551--555.

\bibitem{Zhong:2021epq}
T.~Zhong, Z.-H. Zhu, H.-B. Fu, X.-G. Wu, T.~Huang, {An improved light-cone
  harmonic oscillator model for the pionic leading-twist distribution amplitude
  -- arXiv:2102.03989 [hep-ph]}.

\bibitem{Zhang:2017bzy}
J.-H. Zhang, J.-W. Chen, X.~Ji, L.~Jin, H.-W. Lin, {Pion Distribution Amplitude
  from Lattice QCD}, Phys. Rev. D 95 (2017) 094514.

\bibitem{Chen:2017gck}
J.-H. Zhang, et~al., {Kaon Distribution Amplitude from Lattice QCD and the
  Flavor SU(3) Symmetry}, Nucl. Phys. B 939 (2019) 429--446.

\bibitem{Bali:2019dqc}
G.~S. Bali, V.~M. Braun, S.~B\"urger, M.~G\"ockeler, M.~Gruber, F.~Hutzler,
  P.~Korcyl, A.~Sch\"afer, A.~Sternbeck, P.~Wein, {Light-cone distribution
  amplitudes of pseudoscalar mesons from lattice QCD}, JHEP 08 (2019) 065,
  [Addendum: JHEP 11, 037 (2020)].

\bibitem{Lepage:1979zb}
G.~P. Lepage, S.~J. Brodsky, {Exclusive Processes in Quantum Chromodynamics:
  Evolution Equations for Hadronic Wave Functions and the Form-Factors of
  Mesons}, Phys. Lett. B 87 (1979) 359--365.

\bibitem{Efremov:1979qk}
A.~V. Efremov, A.~V. Radyushkin, {Factorization and Asymptotical Behavior of
  Pion Form- Factor in QCD}, Phys. Lett. B 94 (1980) 245--250.

\bibitem{Lepage:1980fj}
G.~P. Lepage, S.~J. Brodsky, {Exclusive Processes in Perturbative Quantum
  Chromodynamics}, Phys. Rev. D 22 (1980) 2157--2198.

\bibitem{Roberts:2021xnz}
C.~D. Roberts, {On Mass and Matter}, AAPPS Bulletin 31 (2021) 6.

\bibitem{Chang:2011ei}
L.~Chang, C.~D. Roberts, {Tracing masses of ground-state light-quark mesons},
  Phys. Rev. C 85 (2012) 052201(R).

\bibitem{Binosi:2016wcx}
D.~Binosi, L.~Chang, J.~Papavassiliou, S.-X. Qin, C.~D. Roberts, {Natural
  constraints on the gluon-quark vertex}, Phys. Rev. D 95 (2017) 031501(R).

\bibitem{Binosi:2016nme}
D.~Binosi, C.~Mezrag, J.~Papavassiliou, C.~D. Roberts,
  J.~Rodr{\'i}guez-Quintero, {Process-independent strong running coupling},
  Phys. Rev. D 96 (2017) 054026.

\bibitem{Cui:2019dwv}
Z.-F. Cui, J.-L. Zhang, D.~Binosi, F.~de~Soto, C.~Mezrag, J.~Papavassiliou,
  C.~D. Roberts, J.~Rodr{\'{\i}}guez-Quintero, J.~Segovia, S.~Zafeiropoulos,
  {Effective charge from lattice QCD}, Chin. Phys. C 44 (2020) 083102.

\bibitem{Deur:2016tte}
A.~Deur, S.~J. Brodsky, G.~F. de~Teramond, {The QCD Running Coupling}, Prog.
  Part. Nucl. Phys. 90 (2016) 1--74.

\bibitem{Brodsky:2008be}
S.~J. Brodsky, R.~Shrock, {Maximum Wavelength of Confined Quarks and Gluons and
  Properties of Quantum Chromodynamics}, Phys. Lett. B 666 (2008) 95--99.

\bibitem{Aguilar:2015bud}
A.~C. Aguilar, D.~Binosi, J.~Papavassiliou, {The Gluon Mass Generation
  Mechanism: A Concise Primer}, Front. Phys. China 11 (2016) 111203.

\bibitem{Gao:2017uox}
F.~Gao, S.-X. Qin, C.~D. Roberts, J.~Rodr{\'{\i}}guez-Quintero, {Locating the
  Gribov horizon}, Phys. Rev. D 97 (2018) 034010.

\bibitem{Huber:2018ned}
M.~Q. Huber, {Nonperturbative properties of Yang-Mills theories}, Phys. Rept.
  879 (2020) 1 -- 92.

\bibitem{Dokshitzer:1977sg}
Y.~L. Dokshitzer, Calculation of the structure functions for deep inelastic
  scattering and e+ e- annihilation by perturbation theory in quantum
  chromodynamics. ({\mbox {i}n {r}ussian}), Sov. Phys. JETP 46 (1977) 641--653.

\bibitem{Gribov:1972ri}
V.~Gribov, L.~Lipatov, {Deep inelastic e p scattering in perturbation theory},
  Sov. J. Nucl. Phys. 15 (1972) 438--450.

\bibitem{Lipatov:1974qm}
L.~N. Lipatov, {The parton model and perturbation theory}, Sov. J. Nucl. Phys.
  20 (1975) 94--102.

\bibitem{Altarelli:1977zs}
G.~Altarelli, G.~Parisi, {Asymptotic Freedom in Parton Language}, Nucl. Phys. B
  126 (1977) 298--318.

\bibitem{Sufian:2019bol}
R.~S. Sufian, J.~Karpie, C.~Egerer, K.~Orginos, J.-W. Qiu, D.~G. Richards,
  {Pion Valence Quark Distribution from Matrix Element Calculated in Lattice
  QCD}, Phys. Rev. D 99 (2019) 074507.

\bibitem{Hecht:2000xa}
M.~B. Hecht, C.~D. Roberts, S.~M. Schmidt, {Valence-quark distributions in the
  pion}, Phys. Rev. C 63 (2001) 025213.

\bibitem{Grunberg:1982fw}
G.~Grunberg, {Renormalization Scheme Independent QCD and QED: The Method of
  Effective Charges}, Phys. Rev. D 29 (1984) 2315.

\bibitem{Dokshitzer:1998nz}
Y.~L. Dokshitzer, {\emph{Perturbative QCD theory (includes our knowledge of
  \mbox{$\alpha(s)$})} - hep-ph/9812252}, in: {High-energy physics.
  Proceedings, 29th International Conference, ICHEP'98, Vancouver, Canada, July
  23-29, 1998. Vol. 1, 2}, 1998, pp. 305--324.

\bibitem{Joo:2019bzr}
B.~Jo\'o, J.~Karpie, K.~Orginos, A.~V. Radyushkin, D.~G. Richards, R.~S.
  Sufian, S.~Zafeiropoulos, {Pion valence structure from Ioffe-time parton
  pseudodistribution functions}, Phys. Rev. D 100 (2019) 114512.

\bibitem{Gao:2020ito}
X.~Gao, L.~Jin, C.~Kallidonis, N.~Karthik, S.~Mukherjee, P.~Petreczky,
  C.~Shugert, S.~Syritsyn, Y.~Zhao, {Valence parton distribution of the pion
  from lattice QCD: Approaching the continuum limit}, Phys. Rev. D 102~(9)
  (2020) 094513.

\bibitem{patrick_barry_2020_4019411}
P.~Barry, {JAM pion PDF analysis including resummation} (Jun. 2020).
\newblock \href {http://dx.doi.org/10.5281/zenodo.4019411}
  {\path{doi:10.5281/zenodo.4019411}}.

\bibitem{Roberts:2016vyn}
C.~D. Roberts, {Perspective on the origin of hadron masses}, Few Body Syst. 58
  (2017) 5.

\end{thebibliography}

\end{document}